\documentclass[prl,twocolumn,showpacs,floatfix,amsmath,amssymb,superscriptaddress]{revtex4}
\usepackage{latexsym}
\usepackage{graphicx}
\usepackage{times}
\usepackage{color}
\usepackage{amsmath}
\usepackage{dcolumn}
\usepackage{latexsym,amsmath,amssymb,bm,euscript}
\begin{document}

\title{Chern Half Metals: A New Class of Topological Materials to Realize the Quantum Anomalous Hall Effect}

\author{Jun Hu$^*$}
\affiliation{College of Physics, Optoelectronics and Energy, Soochow University, Suzhou, Jiangsu 215006, China}
\affiliation{Department of Physics and Astronomy, University of California, Irvine, California 92697-4575, USA}
\author{Zhenyue Zhu}
\affiliation{Department of Physics and Astronomy, University of California, Irvine, California 92697-4575, USA}
\author{Ruqian Wu$^*$}
\affiliation{Department of Physics and Astronomy, University of California, Irvine, California 92697-4575, USA}


\begin{abstract}
{The search of new topological insulators that demonstrate the quantum anomalous Hall effect (QAHE) is a cutting-edge research topic in condensed matter physics and materials science. So far, the QAHE has been observed only in Cr-doped (Bi,Sb)$_2$Te$_3$ at extremely low temperature. Therefore, it is important to find new materials with large topological band gap and high thermal stability for the realization of the QAHE. Based on first-principles and tight-binding model calculations, we discovered a new class of topological phase---Chern half metal which manifests the QAHE in one spin channel while is metallic in the other spin channel---in Co or Rh deposited graphene. The QAHE is robust in these sytems for the adatom coverage ranging from 2\% to 6\%. Meanwhile, these systems have large perpendicular magnetic anisotropy energies of 5.3 and 11.5 meV, necessary for the observation of the QAHE at reasonably high temperature.}
\end{abstract}


\maketitle

The recent discovery of topological insulators (TIs)---that act as insulators in their bulks yet possess quantized conducting edge or surface states---has triggered extremely active multidisciplinary research activities for the exploration of new fundamental science and exotic materials \cite{KaneMele, HgTeQSHtheory, HgTeQSHexpt, Nagaosa, KaneReview, QiReview}. One of the most interesting phenomena in this realm is the quantum anomalous Hall effect (QAHE) which is the combined consequence of the spin-orbit coupling (SOC) and reduction of the time-reversal symmetry due to \textit{intrinsic} magnetization. These magnetic TIs, also called as Chern insulators since the QAHE is characterized by nonzero integer Chern numbers \cite{ChernNumber, TKNN}, can be produced by using impurities \cite{ZhangQAH, FangQAH}, adatoms \cite{Qiao-Fe, Blugel-5d, Liu-Wu-Ming}, or substrate \cite{Vanderbilt_Pb, Qiao-Proximity} along with regular TIs such as graphene and Bi-based compounds. Furthermore, it is necessary to align their magnetization along the surface normal, in analog to the geometry of conventional anomalous Hall effect \cite{AHE}. Therefore, the realization of the QAHE in experiment is challenging. The only experimental observation of the QAHE was reported recently in Cr-doped (Bi,Sb)$_2$Te$_3$ in an extreme experimental condition below 0.1 K \cite{XueQAH}, due to the tiny TI gap and other complex factors in dealing with surfaces of three-dimensional TIs. Obviously, searching for more feasible Chern insulators especially in simple two-dimensional (2D) systems with large nontrivial band gaps is of great importance for the realization and exploitation of the QAHE.

\begin{figure}[b]
\includegraphics[width = 8.5 cm]{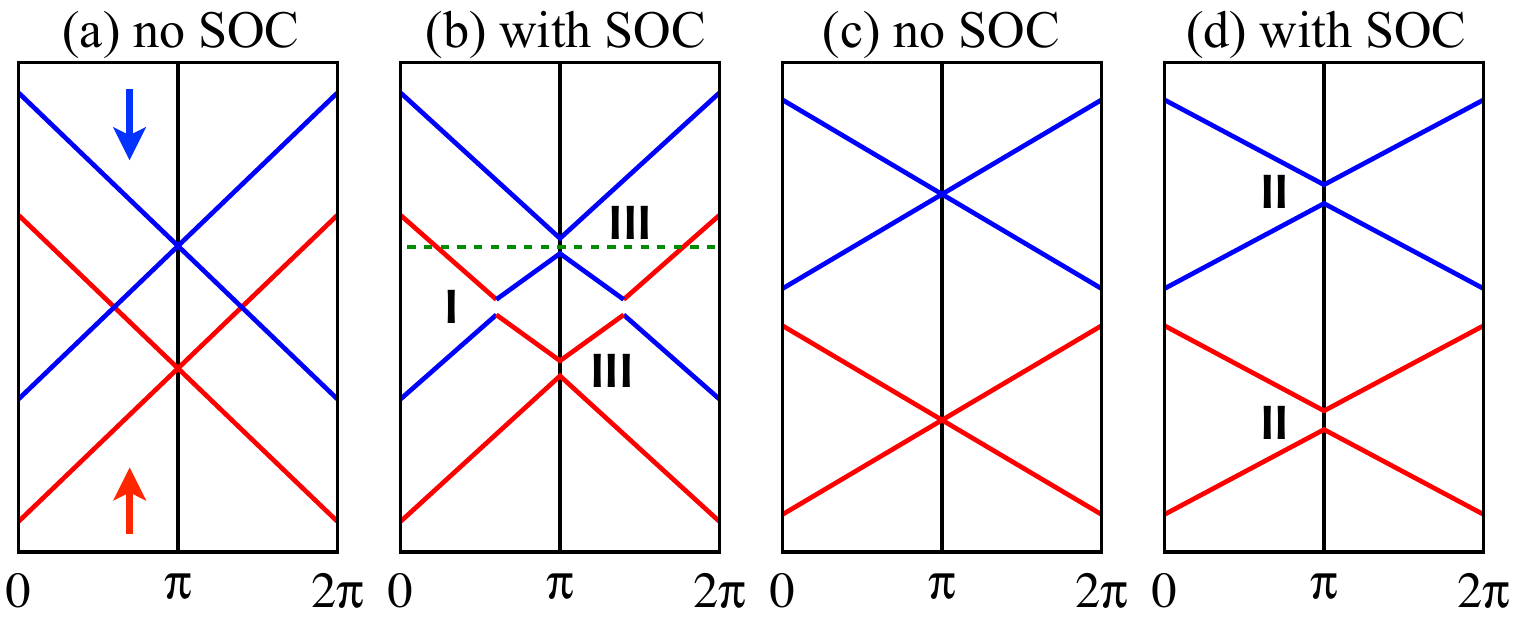}
\caption{Schematic band structures for classifying the topological insulator gaps for the QAHE. (a) and (b) With moderate exchange field,  (c) and (d) with strong exchange field. The red and blue arrows indicate the majority spin and minority spin states, respectively. Three types of SOC induced gaps are marked by `I', `II' and `III'. The green dashed line in (b) indicates the $E_F$ for the type-III gap.
}
\label{TI-gap}
\end{figure}

The TI gaps of Chern insulators reported so far can be sorted into two groups, depending on the strength of \textit{intrinsic} magnetization ($M_{ex}$). As presented by the schematic band structures in Fig. \ref{TI-gap}(a) and \ref{TI-gap}(b), most Chern insulators reported in literature \cite{ZhangQAH, FangQAH, Qiao-Fe, Blugel-5d, XueQAH} have the type-I TI gap which is produced by the interspin SOC interaction under moderate $M_{ex}$. The type-II TI gap which is induced by the intraspin SOC interaction [Fig. \ref{TI-gap}(c) and \ref{TI-gap}(d)] under large $M_{ex}$ was also predicted in an organic material \cite{Liu_Feng_QAH}. Obviously, there can be another type of SOC induced band gap [denoted as type-III in Fig. \ref{TI-gap}(b)], that may lead to a half metallic feature. If the Fermi level ($E_F$) is within the gap, one may expect that the spin channel with the gap behaves like a Chern insulator, whereas the other spin channel is metallic. In this sense, materials with the type-III gap can be classified as {\textit {Chern half metals}} --- a new class of topological materials for spintronics applications.

Graphene is an excellent template to realize the QAHE in 2D systems at high temperature, due to its structural simplicity and stability. Huge TI gaps can be produced in graphene by adatoms, through hybridization with its $\pi$ bands \cite{PRX-Hu-QSH, PRL-Hu-QSH}. The QAHE was predicted in Fe \cite{Qiao-Fe} or W \cite{Blugel-5d} deposited graphene, with type-I TI gaps of 5.5 or $\sim$30 meV. However, the spin orientation of adatoms in both cases lies in the graphene plane (see Fig. S1 and Ref. \cite{Blugel-5d}), rather than perpendicular to it as required by the QAHE. Clearly, it is equally important to determine the magnetic anisotropy energies of potential Chern insulators, for the correct alignment of their magnetization and also for their thermal stability.

In this paper, we use the first-principles and tight-binding (TB) model calculations to demonstrate that the graphene with Co or Rh adatoms can be excellent Chern half metals. Both Co/graphene and Rh/graphene have giant type-III TI gaps (50 and 100 meV, respectively) and nonzero integer Chern number, indicating the existence of the QAHE. The quantized edge states are highly localized within about 2.2 nm from the edge, and robust against randomness of the adatom distributions. Their large magnetic anisotropy energies (5.5 and 11 meV, respectively) can protect the QAHE at temperature up to $\sim$ 60 K.

Since most $3d$ transition metal adatoms on graphene may generate type-I TI gaps \cite{Qiao-Fe, Niu-3d}, we first investigate how many of them can have the perpendicular magnetization, by calculating their magnetic anisotropy energies, $MAE = E(\rightarrow) - E(\uparrow)$, with the torque method \cite{torque1, torque2, torque3}. We found that only Mn (MAE $=+1.4$ meV) and Co (MAE $=+5.3$ meV) adatoms on graphene have perpendicular magnetic anisotropy, in good agreement with recent experimental measurement \cite{Co-perpendicular}. While a type-I TI gap at $\sim$0.8 eV below the $E_F$ is produced in Mn/graphene, a large type-III gap close to the $E_F$ is found in Co/graphene. So it is interesting and important to reveal whether the type-III gap of Co/graphene has topological-insulator feature.

\begin{figure}
\includegraphics[width = 8.5 cm]{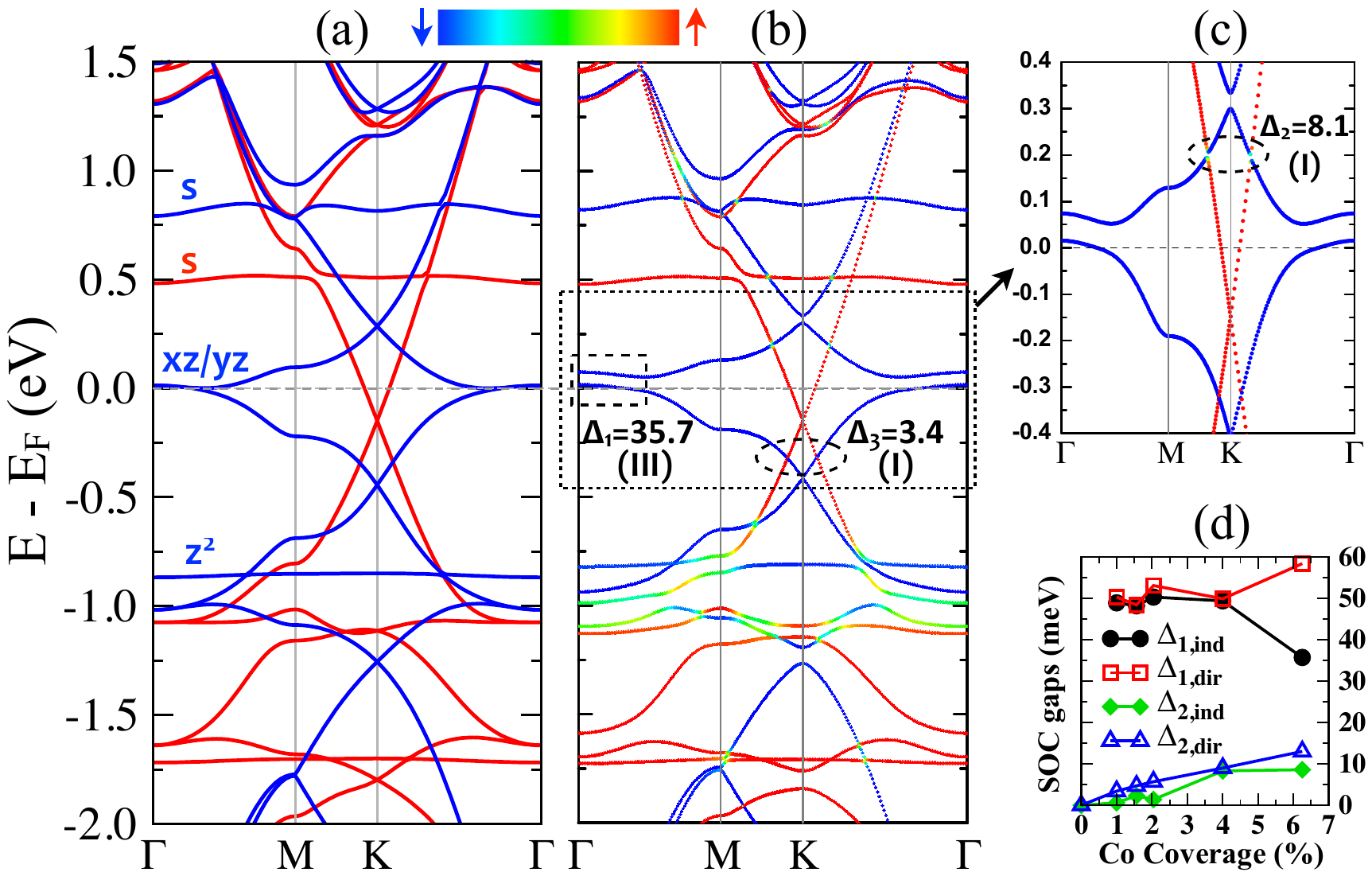}
\caption{Electronic properties of a 4$\times$4 graphene with one Co adatom. (a-c) The spin-resolved band structures without and with the SOC interaction. The horizontal grey dashed line indicates the $E_F$. The color bar indicates the relative ratio of different spins. Three SOC induced gaps are given in meV, with their classifications in the parentheses. (d) The Co coverage dependences of $\mathrm{\Delta_1}$ and $\mathrm{\Delta_2}$. Here both the indirect and direct band gaps are plotted.
}
\label{band-Co}
\end{figure}

To reveal the origin of the type-III gap of Co/graphene, it is useful to discuss its electronic properties without the involvement of SOC. As shown by the spin-resolved band structures in Fig. \ref{band-Co}(a) and projected density of states (PDOS) in Fig. S3(a), the Co-3d orbitals split into three groups: $d_{z^2}$, $d_{xz/yz}$, and $d_{xy/x^2-y^2}$ in the crystal field of graphene. All Co-3d orbitals except one branch of the $d_{xz/yz}$ orbitals in the minority spin channel are fully occupied, resulting in an electronic configuration of $d^9s^0$ and a spin moment of $\sim$ 1.0 $\mu_B$. It should be pointed out that the adsorption geometry and electronic state of Co on graphene are still in debate \cite{Co-on-top, Co-on-hollow}. Nevertheless, it is commonly believed that the Co adatoms take the hollow sites on graphene and adopt the electronic configuration of $d^9s^0$ \cite{Niu-3d, Zhang-Shiwei-Co}, as found in our calculations.  Furthermore, the Coulomb correlation has negligible effect on the electronic structures of Co/graphene, as been discussed in literature for similar systems \cite{Blugel-5d, Liu-Wu-Ming}.

Interestingly, the SOC interaction lifts the degeneracy of the $d_{xz/yz}$ orbitals around the $\Gamma$ point and opens a large type-III band gap ($\Delta_1$) of 35.7 meV (or a direct band gap of $\sim$58 meV), as shown in Fig. \ref{band-Co}(b) and \ref{band-Co}(c). The Fermi level is pinned at $\sim$ 15 meV bellow the gap because of the existence of a tiny electron pocket in the majority spin channel. Clearly, the Fermi level can be tuned easily into this gap by applying a small positive gate voltage. Meanwhile, the majority spin states show linear dispersion near the $E_F$ and are not much affected by the SOC interaction. As a result, Co/graphene is a half metal when the  $E_F$ locates in the gap $\Delta_1$: conducting in the majority spin channel but insulating in the minority spin channel. We found that the gap $\Delta_1$ opens mainly due to the strong SOC effect through $\langle L_z \rangle = \langle d_{xz/yz}^o, \downarrow | L_z | d_{xz/yz}^u,\downarrow \rangle$ of Co \cite{note}. Meanwhile, the magnetic anisotropy is the result of the competition between $\langle L_z \rangle$ and $\langle L_{\pm} \rangle$ \cite{MAE-model, GiantMAE}, so the large matrix element $\langle L_z \rangle$ is also the origin of the strong perpendicular magnetic anisotropy of Co/graphene. Intriguingly, the magnitude of $\Delta_1$ remains almost unchanged even when the coverage is reduced to 1\%, as shown in Fig. \ref{band-Co}(d). We also found two type-I gaps for Co/graphene ($\Delta_2=8.1$ meV and $\Delta_3=3.4$ meV) around $+$0.19 eV and $-$0.32 eV, respectively [Fig. \ref{band-Co}(b) and \ref{band-Co}(c)]. These gaps stem from the interspin SOC interactions, and they decrease monotonically with the drop of the Co coverage [Fig. \ref{band-Co}(d)]. 

\begin{figure}
\includegraphics[width = 8.5 cm]{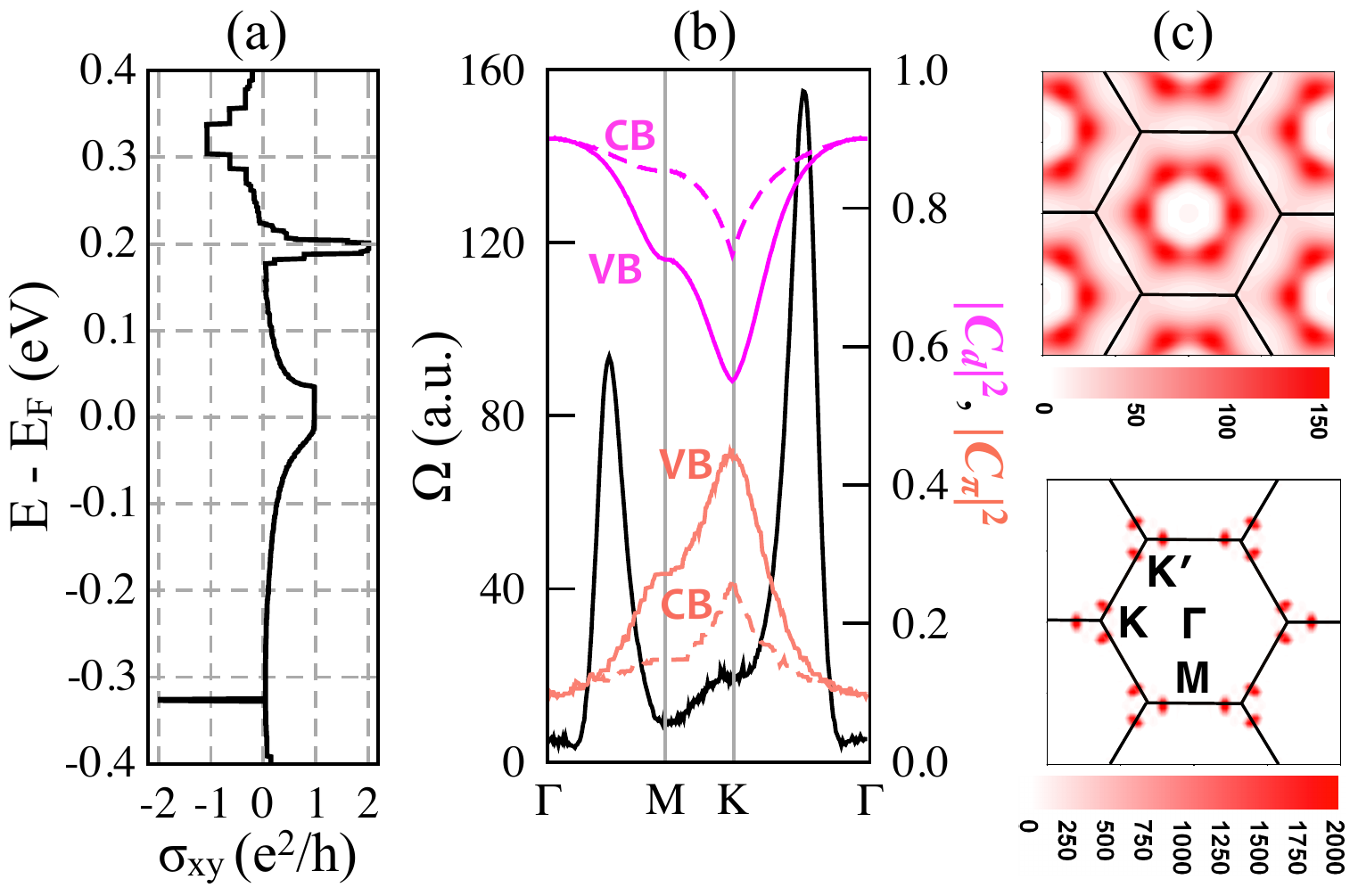}
\caption{(a) Fermi level-dependent anomalous Hall conductance ($\sigma_{xy}$). (b) Berry curvature $\Omega$ (black curve) for the gap $\Delta_1$ and weights of the Co $3d$ (magenta curves) and graphene $\pi$ orbitals (salmon curves) for the highest valence band ({\textbf VB}) and lowest conduction band ({\textbf CB}). (c) Distribution of Berry curvature in the whole 2DBZ for $\Delta_1$ (upper panel) and $\Delta_2$ (lower panel).
}
\label{qahe}
\end{figure}

Now we need to see if the SOC induced gaps are nontrivial and give rise to the QAHE \cite{FangQAH, Qiao-Fe, Blugel-5d}, which is characterized by nonzero Chern number \cite{ChernNumber, TKNN}
\begin{equation}
C=\frac{1}{2\pi}\int_{BZ}\Omega({\textbf k}) d^2k.
\end{equation}
The Berry curvature $\Omega({\textbf k})$ can be determined as \cite{TKNN, Yao-AHE}
\begin{equation}
\Omega({\textbf k}) = 2{\mathrm{Im}} \sum_{n \in \{o\}} \sum_{m \in \{u\}} \frac{\langle\psi_{n{\textbf k}}|v_x|\psi_{m{\textbf k}}\rangle \langle \psi_{m{\textbf k}}|v_y|\psi_{n{\textbf k}}\rangle}{(\varepsilon_{m{\textbf k}}-\varepsilon_{n{\textbf k}})^2},
\end{equation}
where $\{o\}$ and $\{u\}$ stand for the sets of occupied and unoccupied states, respectively; $\psi_{n{\textbf k}}$ and $\varepsilon_{n{\textbf k}}$ are the spinor Bloch wavefunction and eigenvalue of the $n$th band at {\textbf k} point; and $v_{x(y)}$ is the velocity operator. 
The $E_F$ dependent Hall conductance: $\sigma_{xy}=C(e^2/h)$ of Co/graphene is plotted in Fig. \ref{qahe}(a). It is obvious that $\sigma_{xy}$ is quantized in all gaps mentioned above. Since $\Delta_2$ and $\Delta_3$ are type-I TI gaps are similar to what was reported for Fe/graphene \cite{Qiao-Fe}, it is understandable that they have $C=\pm2$. One important finding here is $C=1$ in the type-III gap $\Delta_1$, even though the majority spin channel is metallic. We also calculated the spin-resolved Chern numbers and found that the contribution of the majority spin states to the Chern number within $\Delta_1$ is negligible. Therefore, the quantized $\sigma_{xy}$ within $\Delta_1$ results dominantly from the minority spin states. We may perceive that a Co/graphene nanoribbon has quantized current carrying minority spin along the edges, together with majority spin current in the interior region. This may offer new opportunities for the design of spin filters and spintronics devices.

To explore the origin of the topological properties of Co/graphene, we plot the $\Omega({\textbf k})$ along the high symmetric direction in the two-dimensional Brillouin zone (2DBZ) in Fig. \ref{qahe}(a). There are two pronounced peaks around the middle points of $\overline{\bf{\Gamma}{K}}$ and $\overline{\bf{\Gamma M}}$. From the velocity matrix elments $\langle v_{x(y)} \rangle$, we found that the coupling between the highest valence band and lowest conduction band in the minority spin channel (refered as `VB' and `CB') plays the dominant role in $\Omega({\textbf k})$. The nonzero $\langle\psi_{CB,{\textbf k}}|v_x|\psi_{VB,{\textbf k}}\rangle$ requires selection rules of $\Delta l = \pm 1$ and $\Delta m = \pm 1$. Accordingly, the $\Omega({\textbf k})$ must be from the hybridization between the Co $d_{xz/yz}$ and graphene $\pi$ states.  If we decouple the wavefunctions of VB and CB as $\psi_{n{\textbf k}}=C_d^n \phi_d+C^n_{\pi} \phi_{\pi}$, we found that the weight of the $\pi$ states $|C_{\pi}|^2$ is very small around the $\Gamma$ point and increases significantly near the middle points of $\overline{\bf{\Gamma}{K}}$ and $\overline{\bf{\Gamma M}}$ as shown in Fig. \ref{qahe}(b). On the other hand, the energy separation between these two bands increases rapidly beyond the middle points of $\overline{\bf{\Gamma}{K}}$ and $\overline{\bf{\Gamma M}}$, which leads to an abrupt decrease of $\Omega({\textbf k})$. More complete information regarding the distribution of $\Omega({\textbf k})$ in the whole 2DBZ is displayed in the upper panel of Fig. \ref{qahe}(c). For comparison, we also plotted the distribution of the $\Omega({\textbf k})$ of the gap $\Delta_2$ in the lower panel of Fig. \ref{qahe}(c). It can be seen that large areas in the BZ have contributions to $\Omega({\textbf k})$ of $\Delta_1$, but only very tiny regions around the K and K$^\prime$ points where bands cross have contribution to $\Omega({\textbf k})$ of $\Delta_2$.

\begin{figure}
\includegraphics[width = 8.5 cm]{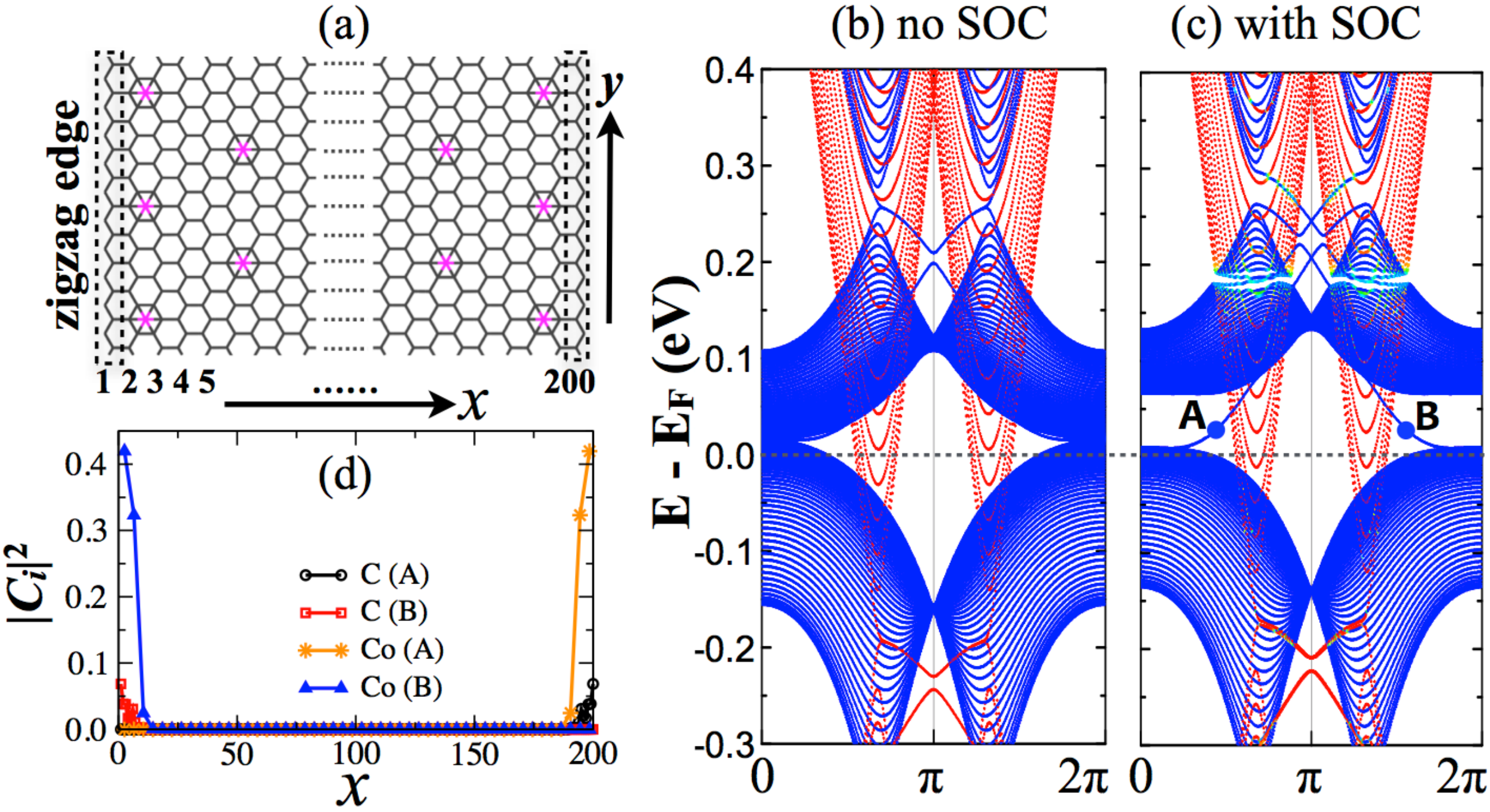}
\caption{(a) The geometry of the the Co/graphene nanoribbon used in our calculation. The width of the nanoribbon is 200 zigzag C chains, and the Co adatoms are distributed uniformly with coverage of 6.25\% (same as one Co adatom in $4\times4$ graphene). The open and periodic directions are notated as {\textit x} and {\textit y}, respectively. (b) and (c) The band structures of Co/graphene nanoribbon without and with SOC. The gray dashed line indicates the natural $E_F$. The color scheme is the same as that in Fig. \ref{band-Co}. (d) The spatial distributions of the edge states A and B as marked in (c). $|C_i|^2$ is the weights of the zigzag C chains and Co adatoms.
}
\label{TB}
\end{figure}

It is known that nonzero integer Chern numbers for the type-I and type-II TI gaps of an infinite 2D material guarantee the existence of quantized edge states when the material is cut into one-dimensional ribbons \cite{ChernNumber}. To check whether similar quantized edge states exist for the type-III gap $\Delta_1$, we developed a TB model with the Hamiltonian $H=H_g+H_a+H_c$ \cite{PRL-Hu-QSH} to directly study the electronic structure of a Co/graphene nanoribbon (NR). Here, $H_g$, $H_a$ and $H_c$ describe graphene, Co, and the hopping between Co 3d orbitals and graphene $\pi$ orbitals, respectively. The details about our TB Hamiltonian can be found in the Supplemental Material. The parameters were obtained by fitting the DFT band structures of the $4\times4$ Co/graphene in Fig. \ref{band-Co}(a) and \ref{band-Co}(b). The TB band structures (Fig. S5) agree with the DFT band structures very well, which demonstrates the high quality of our approach for the description of Co/graphene.

We used a graphene NR with zigzag edges on both sides [Fig. \ref{TB}(a)] to show the existence of the quantized edge states. The width of the graphene NR is 200 zigzag C chains, and the periodicity along the $y$-axis is 4 times of that of graphene. The Co adatoms with a coverage of 6.25\% are distributed uniformly on the graphene NR. The corresponding band structures without and with SOC in the one-dimensional Brillouin zone (1DBZ) are plotted in Figs. \ref{TB}(b) and \ref{TB}(c). It can be seen that the $d_{xz/yz}$ bands near the $E_F$ remain close without SOC. When SOC is invoked, they are separated by about 54 meV, slightly smaller than the gap $\Delta_1$ (58 meV) of the 2D Co/graphene [Fig. \ref{TB}(c) and Fig. S5]. Interestingly, we can see that two gapless bands cross the bulk gap $\Delta_1$, presumably being from the edges. To prove this, we calculated the weights of all atoms for the wavefunctions of the states A and B as marked in Fig. \ref{TB}(c). The summations over atoms with the same $x$ position are plotted in Fig. \ref{TB}(d). Clearly, their wavefunctions indeed localize near the right and left edges, respectively, and the weights decay very rapidly to zero towards the center of the Co/graphene NR. The depth of the edge states is about 22 {\AA}. We want to point out that the edge states do not interact with the majority spin states even they cross each other in the gap $\Delta_1$, as presented in Fig. \ref{TB}(c).

When the Fermi level locates in the gap $\Delta_1$, the edge states contribute to the quantized conductance. Because the velocity of conducting electrons can be calculated as $\vec{v}(\textbf{k})=\nabla E(\textbf{k})$, the electrons on the right edge move along the $y$ direction, while the electrons on the left edge move oppositely. Moreover, there is only one conducting channel at each edge, corresponding to the Chern number $C=1$ as discussed above.

\begin{figure}
\includegraphics[width = 8.5 cm]{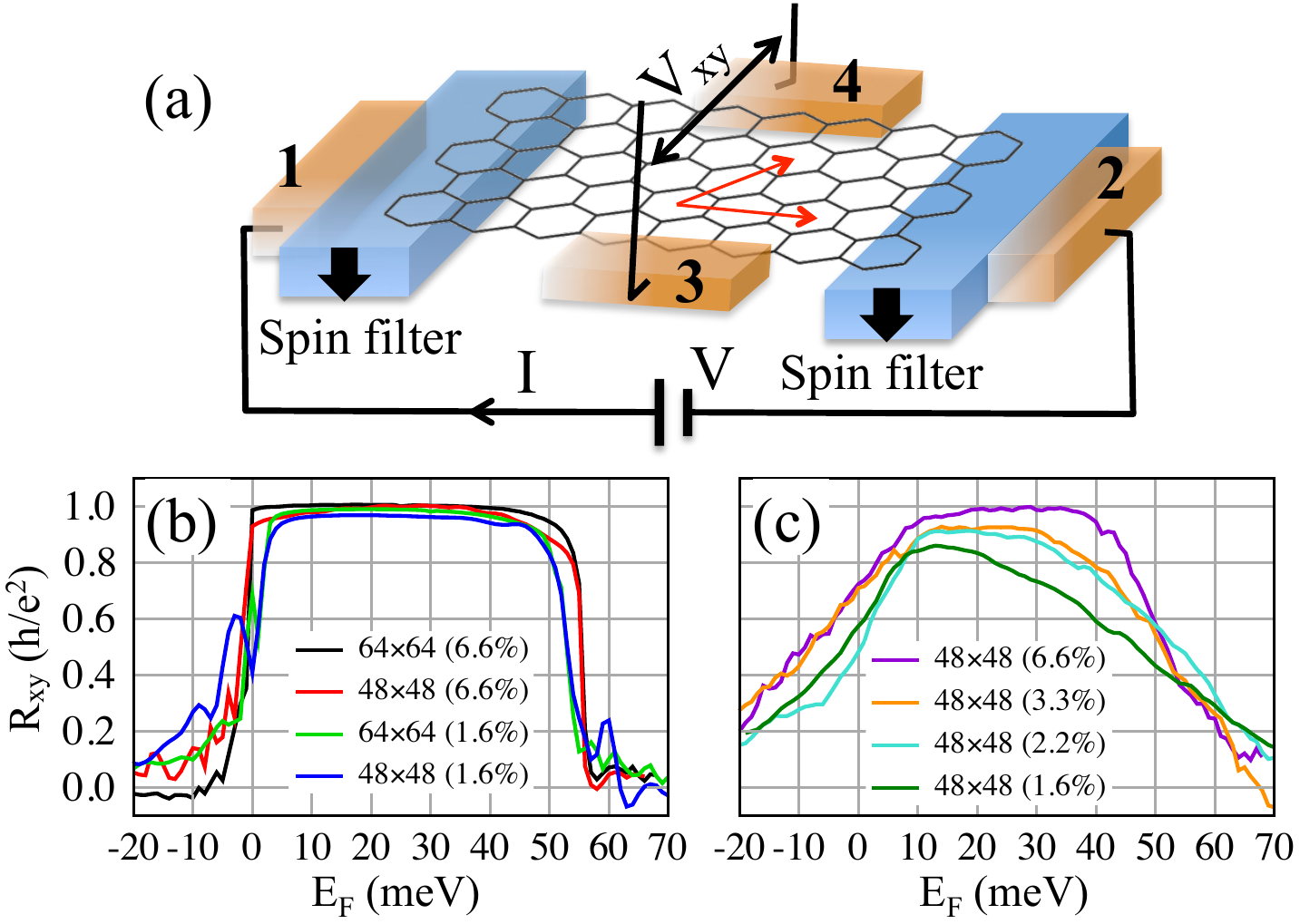}
\caption{(a) Schematic setup to measure the anomalous Hall resistance ($R_{xy}$). The numerals indicate the leads (orange bars). The black fat arrows indicate the magnetization of spin filters (light blue bars). The red arrows show the definition of the size graphene nanosheet (the numbers of hollow sites along the corresponding direction). (b) and (c) $R_{xy}$ of Co/graphene nanosheet for both uniform and random distributions of Co adatoms, respectively.
}
\label{Rxy}
\end{figure}

We also directly calculated the energy-dependent anomalous Hall resistance ($R_{xy}$) of Co/graphene with the Landauer-Buttiker formula \cite{lbf1, lbf2, lbf3}  in a setup shown in Fig. \ref{Rxy}(a).  Although it appears to be difficult to detect the topological edge states in Chern half metals with one spin channel being metallic, the simplicity of the two-dimensional geometry, the large topological band gap, and the high magnetic anisotropy energy we discussed here should be attractive for fundamental studies and device applications. In particular, one may use spin filter materials for the leads in Fig. \ref{Rxy}(a), such as ferromagnetic half metal SrRuO$_4$, so that only the minority spin current flows through the graphene ribbon. The voltage drop $V_{xy}$ on the two edges can be measured across leads 3 and 4, and $R_{xy}$ is defined as $R_{xy}=V_{xy}/I$ (in the unit of $h/e^2$). Details of these calculations are shown in the Supplemental Material. From the $E_F$-dependent $R_{xy}$ in Fig. \ref{Rxy}(b), we can see plateaus above the $E_F$ which correspond to the topological edge states. For the $64\times64$ graphene nanosheet with Co coverage of 6.6\%, we found that $R_{xy} $ may approach to 1 for $E_F$ varying from 0 to 40 meV. For other cases with smaller graphene ribbon or lower Co coverage, $R_{xy}$ still remains larger than 0.97.

We should point out that the edge states in the minority spin channel are well separated from the bulk states in the majority spin channel, as shown in Fig. \ref{TB}(c). However, states of different spins are expected to be mixed in realistic cases where Co adatoms are randomly distributed, which may ruin the quantization of the  $R_{xy}$. To see if the QAHE is robust, different random distribution patterns were calculated for a $48\times48$ graphene nanosheet with Co coverages of 1.6\%, 2.2\%, 3.3\% and 6.6\%, and the averaged values of $R_{xy}$ are plotted in Fig. \ref{Rxy}(c). It is important that the curves of $R_{xy}$ show plateaus even as the coverage of Co is only about 2\%. The peak value of $R_{xy}$ reaches 0.99 for the Co coverage of 6.6\%, indicating the robustness of the QAHE in this case against the randomization of Co adatoms. Moreover, the deviation of $R_{xy}$ from the theoretical value (i.e. $h^2/e$) for the lower-coverage cases should mainly come from the finite size effect in our numerical calculations. Therefore we believe that the QAHE plateau is observable, as long as the the coverage is between $2\%\sim6\%$ and the size of graphene sheet is large enough. This coverage range is estimated based on the requirements of adequately large $\Delta_1$ yet negligible direct Co-Co interaction for the realization of QAHE.

To estimate the temperature limit for the observation of the QAHE, we need consider a few factors. (i) The MAE of 5.3 meV ensures the perpendicular spin orientation stable up to $\sim$ 60 K. (ii) The TI gap (indirect/direct band gap of 35.7/58 meV) in minority spin channel is larger than the exciton energy at room temperature. (iii) The energy barrier of Co adatom diffusion on graphene is about 0.4 eV \cite{TM-adatom-on-graphene}, which implies that the aggregation of Co adatoms can be avoided below 100 K. In fact, recent experiment revealed that over 90\% Co adatoms on graphene exist as monomer \cite{Co-perpendicular}. Therefore, the temperature limit to observe the QAHE in Co/graphene in experiment mainly depends on the stability of the perpendicular spin orientation, i.e. $\sim$ 60 K which is much higher than that in Cr-doped (Bi,Sb)$_2$Te$_3$ \cite{XueQAH}.

The TI gap and MAE may be further enhanced if we use Rh because of its stronger SOC strength. We found that the band structures of Rh/graphene (Fig. S4) are very similar with those of Co/graphene, except that the $E_F$ locates at 20 meV above the conduction band minimum in the minority spin channel, due to the smaller exchange splitting ($M_S=0.8~\mu_B$). Obviously, the size of the type-III gap $\Delta_1$ is much larger than that of Co/graphene, with the indirect (direct) gap of 100 (118) meV. A Chern number of $C=1$ is obtained for $\Delta_1$ of Rh/graphene, which is thus also a Chern half metal and should have more robust topological features than Co/graphene. Furthermore, the MAE of Rh/graphene is 11.5 meV, which indicates the perpendicular spin orientation to be stable up to $\sim$ 140 K.

In summary, Co/graphene and Rh/graphene are revealed to be excellent Chern half metals, through DFT calculations and TB model analyses. We found that the Co/graphene and Rh/graphene have huge type-III TI gaps (about 50 and 100 meV) and large perpendicular magnetic anisotropy energies (5.3 and 11.5 meV). Therefore, they are ideal model systems for the observation of the QAHE at elevated temperature. We expect that the TI phase found here may be generalized to other materials. The pure and dissipationless spin current at the edges of Chern half metals may find significant applications in spintronics and quantum computating.

\section*{ASSOCIATED CONTENT}
\subsection*{Supporting Information}
Description of methods, discussions on the electronic and magnetic properties of Fe, Co and Rh adatoms on graphene, the details about the tight-binding model and transport calculations Co/graphene. This material is available free of charge via the Internet at http://pubs.acs.org.

\section*{AUTHOR INFORMATION}
\subsection*{Corresponding Authors}
$^*$E-mail: jhu@suda.edu.cn (J.H.).

$^*$E-mail: wur@uci.edu (R.W.).

\subsection*{Notes}
The authors declare no competing financial interest.

\section*{ACKNOWLEDGMENTS}
J.H. and Z.Z. thank useful discussions with Jason Alicea and Xiaoliang Qi. This work was supported by DOE-BES Grant No. DE-FG02-05ER46237 and by NERSC for computing time (J.H. and R.W.), NSF Grant No. DMR-1161348 (Z.Z.).

\section*{REFERENCES}

\begin{thebibliography}{29}
\bibitem{note} Here $L_z$ is the $z$ component of the orbital angular momentum; the superscripts $o$ and $u$ stand for the occupied and unoccupied status, respectively; and the arrow $\downarrow$ indicates the minority spin.
\end{thebibliography}

\newpage

\end{document}